# DLSOM: A Deep learning-based strategy for liver cancer subtyping

Fabio Zamio

Department of Computer Science, University of Ilocos Philippines, Ilocos Sur, Philippines

## Abstract

Liver cancer is a leading cause of cancer-related mortality worldwide, with high genetic heterogeneity posing significant challenges to its diagnosis and treatment. While previous studies have primarily relied on limited subsets of genomic data or feature selection techniques to identify liver cancer subtypes, these methods often result in loss of valuable information. In this study, we present DLSOM a novel deep learning framework employing stacked autoencoders to analyze the complete somatic mutation landscape of 1,139 liver cancer samples, encompassing 20,356 protein-coding genes. This approach transforms high-dimensional mutation data into three low-dimensional features, retaining the full discriminatory information required for robust clustering. Using this framework, we identified five distinct liver cancer subtypes (SC1–SC5) with unique mutational, functional, and biological characteristics. Subtypes SC1 and SC2 exhibited significantly higher mutational loads compared to the others, while SC3 displayed the lowest burden, highlighting the heterogeneity of mutational processes. Subtype-specific mutational signatures and trinucleotide motifs further revealed distinct molecular mechanisms underlying each subtype, with several novel and COSMIC-associated signatures identified, including those linked to hypermutation and chemotherapy resistance. Gene ontology and pathway analyses demonstrated unique functional enrichments for each subtype, supporting the biological relevance of our findings. This study provides a comprehensive framework for liver cancer subtype identification, leveraging the full mutational landscape to reveal novel insights into the molecular heterogeneity of the disease. The findings hold significant promise for advancing precision medicine in liver cancer by facilitating the development of subtype-specific diagnostic tools, biomarkers, and targeted therapies. Our approach highlights the potential of deep learning in high-throughput genomic research, offering a scalable solution for addressing cancer complexity.

## Introduction

Liver Cancer is the second most fatal cancer among men after Lung cancer in developed countries [1]. According to the American Cancer Society estimation for 2019 in the United State of America, over 42,030 new cases (29,480 in men and 12,550 in women) will be diagnosed and also about 31,780 patients (21,600 men and 10,180 women) will die of the disease. Liver cancer is also the leading cause of cancerous death worldwide by causing 600,000 deaths each year [2]. The main causes of liver cancer are not yet very clear [3]. However, liver cancer is found to be associated with birth defects, chronic infection with diseases like hepatitis B and C, hemochromatosis, and cirrhosis, and also drinking alcohol [4-6]. For the aim of helping treatment procedure, it is important to be able to identify different subtypes of this cancer with respect to their severity or their symptoms. During the past decades, several studies have been conducted to address this problem [7-11].

One of the main methods for liver cancer subtype identification is to utilize gene expression patterns as input features to train clustering models be able to tackle this problem. Studies relying on this approach mainly consent on two subtypes for this cancer which in fact does not provide a very accurate and distinguishable categorization. Hence, several studies have tried to incorporate other sorts of information to provide better subtype identification. For example, [11] utilized EpCAM-coexpressed signature as well as gene expression in order to classify the human hepatocellular carcinoma (HCC), which is the most common type of primary liver cancer. As a result, they identified two different subtypes namely, EpCAM+ and EpCAM- for liver cancer. Another study, [7] used genome-wide expression profiling in order to identify subtypes based on the regulation of interferon (IFN)-dependent genes. They also utilized the downregulation of apoptosis-relevant genes as their features and similarly identified two subtypes for liver cancer. Later on [9] utilized gene expressions and investigated any difference in gene expression of HCC patients. They also identified two subtypes for the HCC. They have also performed a study on the limited number of genes. The expression profile of these genes indicated precisely the survival time of the HCC patients of these clusters.

Besides gene expression, other various types of omics data such as mRNA, copy number variation, DNA methylation, and mRNA gene expression have been utilized in computational methods to tackle subtype identification in liver cancer [8, 10]. Among them, [10] have utilized miRNA, mRNA, DNA methylation, copy number variation, and mRNA gene expression and integrated them in order to identify five molecular subtypes using a benchmark consisting of 256 patients from The Cancer Genome Atlas (TCGA). In another study, [8] utilized mRNA expression, miRNA expression, CpG methylation, and clinical information to identify subtypes of HCC which comprises the main part of the patients who have liver cancer. In this work, they utilized deep learning (DL) computational framework on multi-omics HCC dataset using a benchmark consisting of 360 patients from the TCGA to identify two subtypes.

Recently, there has been growing interest among researchers in understanding the role of genomic variants in complex diseases, including neurocognitive disorders [12-14] and various cancers [15-20]. This research effort reflects the broader recognition of how genetic variations contribute to disease mechanisms, progression, and treatment response. In particular, advancements in sequencing technologies have made it possible to study the entire genome at unprecedented resolution, enabling researchers to investigate the significance of mutations, structural variations, and regulatory elements in disease development. Several studies have focused on employing deep learning approaches to analyze these genomic variants, leveraging their ability to handle large, complex datasets and uncover hidden patterns that might be missed by traditional methods. For example, studies on neurocognitive disorders have utilized deep learning to analyze copy number variations [11, 20, 21], chromatin interactions, and gene expression profiles, shedding light on the molecular underpinnings of these conditions. Similarly, in other diseases, deep learning has been applied to identify driver mutations, classify cancer subtypes [9, 22, 23], identify enhancers [24] and predict patient outcomes.

Nowadays, due to decreasing the sequencing costs which were significantly time-consuming, there are lots of researchers have focused their studies to assay genomic information. Since there are various types of simple somatic point mutations have been found as one of the major risk factors of cancers and patients of specific cancer have the different molecular mechanism, subtype identification regarding simple somatic point mutations can significantly help treatment procedure. Hence, our investigations in this study have been centered about simple somatic mutations and their impacts on different subtypes in liver cancer. In previous studies, these types of mutations have been always considered as an important part of investigation but the issue with those investigations and work was they have always used some fractions of these mutations as their features for studying on cancers, etc. so they utilized some kind of feature selection in order to reduce the huge volume of data. Hence, this procedure led them to lose some information during the procedure. To tackle this issue, we wanted to provide some kind of procedure in order to involve all of the data and mutations to utilized complete set of data and information. So we have used deep neural network and specially stacked autoencoder in order to transform all of the mutations to small volume of data. In this way, not only we have reduce the huge amount of data regarding to huge number of mutations in each sample, but also we did not lose any information in this procedure. The other issue with the previous works, was they have used different types of feature selection procedure for their work. Hence, different feature selection method provide different features and each of them provide some kind of different information. Due to the transformation of all the mutations, we solve this deficiency and involved all of the information.

In this study, we present DLSOM, a Stacked Auto Encoder (SAE) deep neural network model to incorporate the mutational load from all the available genes in a low dimensional input feature group. We transform all the input features into a feature group consisting of just three features for liver cancer subtype identification. Hence, unlike previous studies that filter out some of the data in order to reduce the input dimension, we incorporate all those features by feature transformation rather than feature reduction. Hence, we get the benefit of all available discriminatory information instead of discarding them. Here we also use SAE [25] and modified its algorithm in order to integrate heterogeneous data [19, 26]. SAE has also applied for analyzing high-dimensional types of data such as gene expression data [18, 27]  and its performance was efficient for those works. We then use the identified features as the input for K-means clustering model and identify five subtypes in liver cancer with strong biological characterization and statistical validation.

## Results

### Deep learning technique for feature transformation

In our pipeline, we used deep learning – autoencoder-- method in order to transform all of the SSMs in protein coding genes to use transformed features for clustering procedure. All of the samples we have in our benchmark have information of their SSMs in all of the their protein-coding genes. Hence, we counted number of SSMs which occurred in each protein coding gene and use that number as an input feature for our neural net network. So, we have 20,159 features, each of them stands for the counted number of the SSMs which occurred in each genes. Due to the huge amount of data we had, obviously we could not use them for clustering procedure. For transforming this information (1,139 samples, 2,159 features for each)  we fed them into an stacked autoencoder. Then, we extracted the 3 transformed features from the bottleneck of this  model (see the detail in the method section).

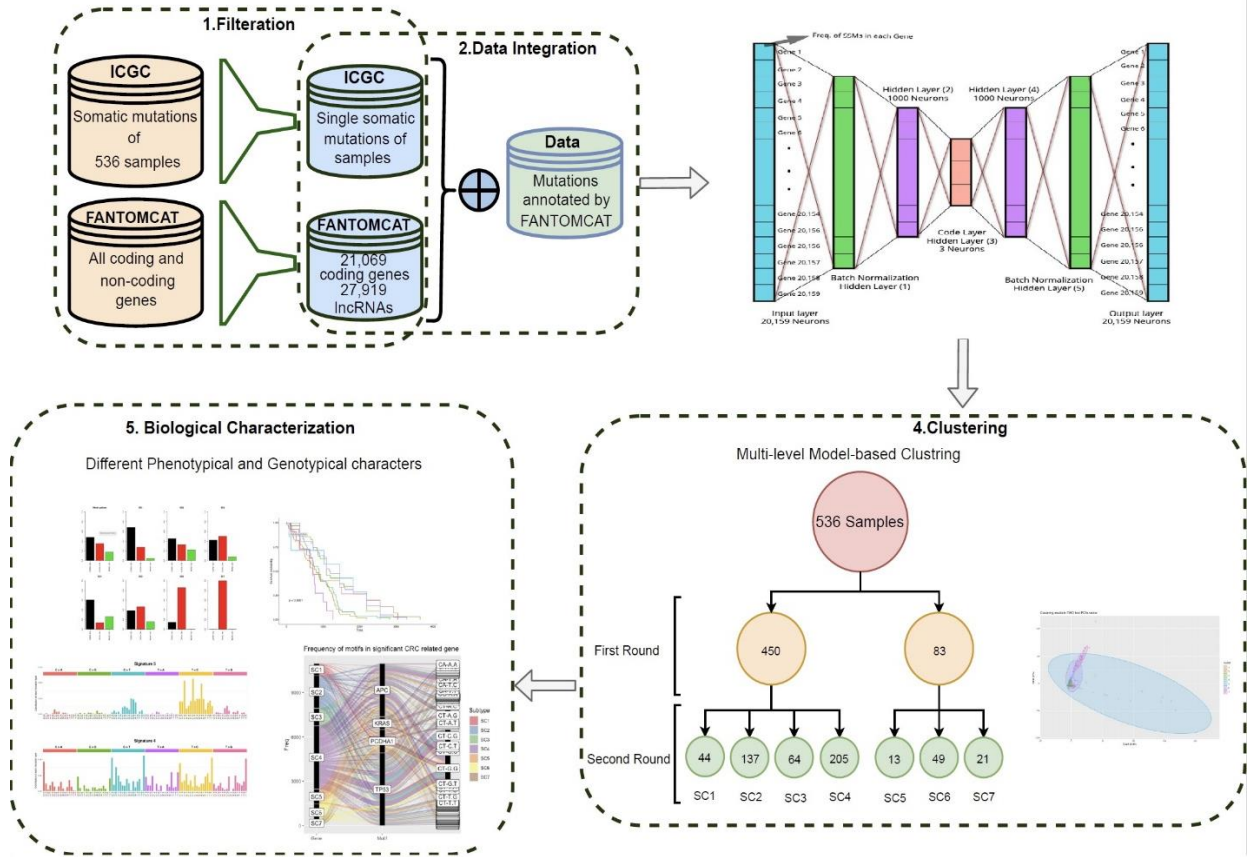

**Figure1.** Conceptual figure of the proposed methods.

**Identify subtypes using extracted features**

Here, we utilized an autoencoder deep learning framework to extract a feature vector consisting of three features which have been transformed using autoencore deep neural network from the whole SSMs in our benchmark. This allows us to have the impact of all coding genes in subtypes identification. This deep learning framework is described in detail in the method section.

After transformation procedure we had three features representing each of our samples. We then used K-means method [28] to cluster liver samples in terms of their transformed features. K-means clustering has been used for various types of tasks in bioinformatics [29-31]. K-means is one of the most prevalent clustering methods which can be combined with the other methods in order to find optimum number of clusters. However, to obtain clusters with K-means method it is required to pass the number of the clusters as a parameter to this algorithm. One of the most challenging issues we have been encountered was to find the optimum number of clusters to pass into the K-means method. To address this issue we have used R package "NbClust" [32]. This R Package contains 30 different statistical methods utilized to find the optimum number of clusters for ia given dataset. In other words, we avoid using any default knowledge and use optimization to identify the number of clusters (*see more details in method section*).

Using NBClust package and K-means clustering, we clustered our samples into five subtypes in the following manner. In the first step, NbClust determines two large clusters for our data regarding 14 indexes (the term index in this package stands for statistical methods which assay the performance of clustering, see more details in method section). It has been confirmed by NbClust that two clusters are the optimum number of clustering. The clusters comprised 329 and 810 samples. Subsequently, in order to provide more insight and more in-depth clustering, we perform another level of clustering for our two main clusters to see if there are any other meaningful sub-clusters that can be produced. Performing Nbclust on each cluster, divides two clusters into five smaller and distinguishable clusters (two for the first cluster and three for the second one, totally 5 five clusters) for liver cancer.

Note that further use of Nbclust and K-means on the identified sub-clusters did not produce any more meaningful and distinguishable clusters (statistical methods illustrate plunge in rating the clusters evaluation) which suggest that the number of meaningful subtypes converges at five. The number of samples for our five subtypes are 152, 172, 780, 20, and 10 samples, respectively. We will refer to these subtypes as SC1, SC2, SC3, SC4, and SC5, respectively. For the rest of this paper we will present the results achieved from different analyses that demonstrates the significance of this study in suggesting a new distinguishable, functionally, and genotypically set of subtypes for liver cancer, which may be used for further drug design to cure liver cancer.

**Mutational rates of each cluster**

We investigated the mutational load for all of the 20,159 protein-coding genes in each cluster by counting number of samples, which have SSM in each protein-coding gene.

Our aim is to compare mutational rate between our identified clusters. As it is shown in **Figure 2** the two larger primer subtypes (S1 and S2) produced as the result of first level of clustering are completely separated with respect to their mutational load. It is also valid for the five subtypes that are produced as the second level of clustering except for subtypes S1 and S2. That separation is not as significant as the results of other subtypes. However, our further analysis that are represented in the next sections clearly demonstrates the significant difference between these two subtypes as well. These two subtypes have significantly higher mutational load than other three subtypes. As **Figure 2** shows, subtypes S3, S4, and S5 are totally separated from each other regarding to their mutational load (more than 70% dissimilarity). **Figure 2** also shows that subtype S3 has low mutational load at protein-coding genes (in average 93 mutation per gene) compared to the rest of subtypes.

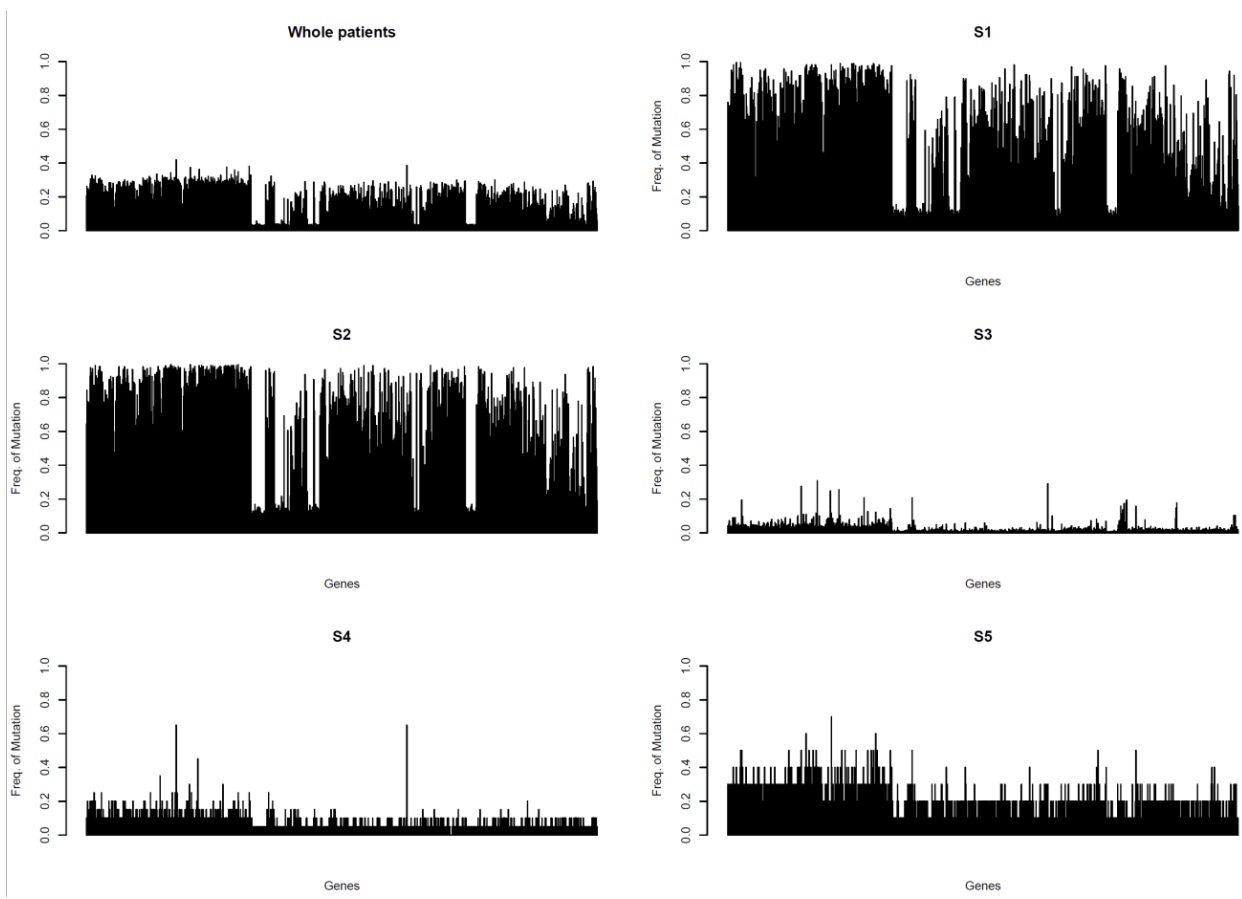

**Figure 2.** Mutational load of each cluster. X-axis is all genes, and Y-axis is the mutational frequency of each gene in the cluster.

### *Most associated genes in each subtype*

After the clustering procedure, we have utilized Fisher's exact test to identify those genes that significantly mutated in each subtype to detect subtype-specific candidate genes. **Figure 3** shows the overlap between most significant genes (top 100 significant genes) in each subtype. As the figure shows for both subtypes SC1 and SC2 there are 28 unique genes that are associated in each individual subtype. Besides, there are also 73 genes that are significantly associated to both of SC1 and SC2 subtypes. The number of unique genes associated with other three subtypes (SC3, SC4, and SC5) are even more significant. Each of the SC3, SC4, and SC5 subtypes has 98 genes, in which they are uniquely associated with them with no overlap with any other subtypes. In the following, we discuss some of the most significantly associated genes in each subtype. Considering their association, all of these genes may be considered as the biomarker and potential target to design related treatment.

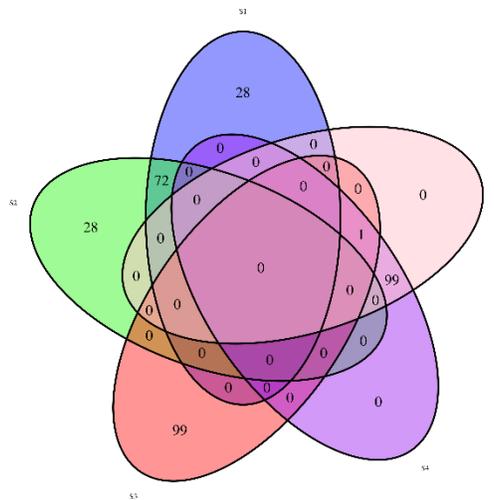

**Figure 3. Venn diagram of common and unique genes between each subtype.** There is zero common genes between all the subtypes.

### *Differential analysis of mutational load in coding and non-coding genes in the liver subtypes*

Different to the previous two sections here we investigated the overall mutational load of coding genes between liver subtypes. To do this, we have subtracted all of the mutational loads of same coding genes in each two pair of the subtypes. This enables us to identify the differences between the mutational load of coding genes in each pair of subtypes. We then picked the genes which caused peaked differences (for which about 15% or higher difference can be found for the pair of subtypes) for each subtype to be investigated if they are related to liver cancer. In The following, we discussed the difference between each two pairs of the combinations, The term difference defined as how much differences can be found between the percentage of the mutational load of a specific gene between two subtypes. **Figure 4** presents a comparative analysis of the mutation load between protein-coding and non-coding genes in liver cancer. Figure 4a illustrates the mutation load across 20,356 protein-coding genes, while Figure 4b (b) focuses on 16,195 long non-coding RNAs (lncRNAs). Each principal component score (PCS) highlights distinct mutation distribution patterns, demonstrating variation in mutation intensity across different genomic regions. The results provide insights into how mutations are differentially distributed between coding and non-coding regions, underscoring the potential importance of non-coding elements in liver cancer pathogenesis.

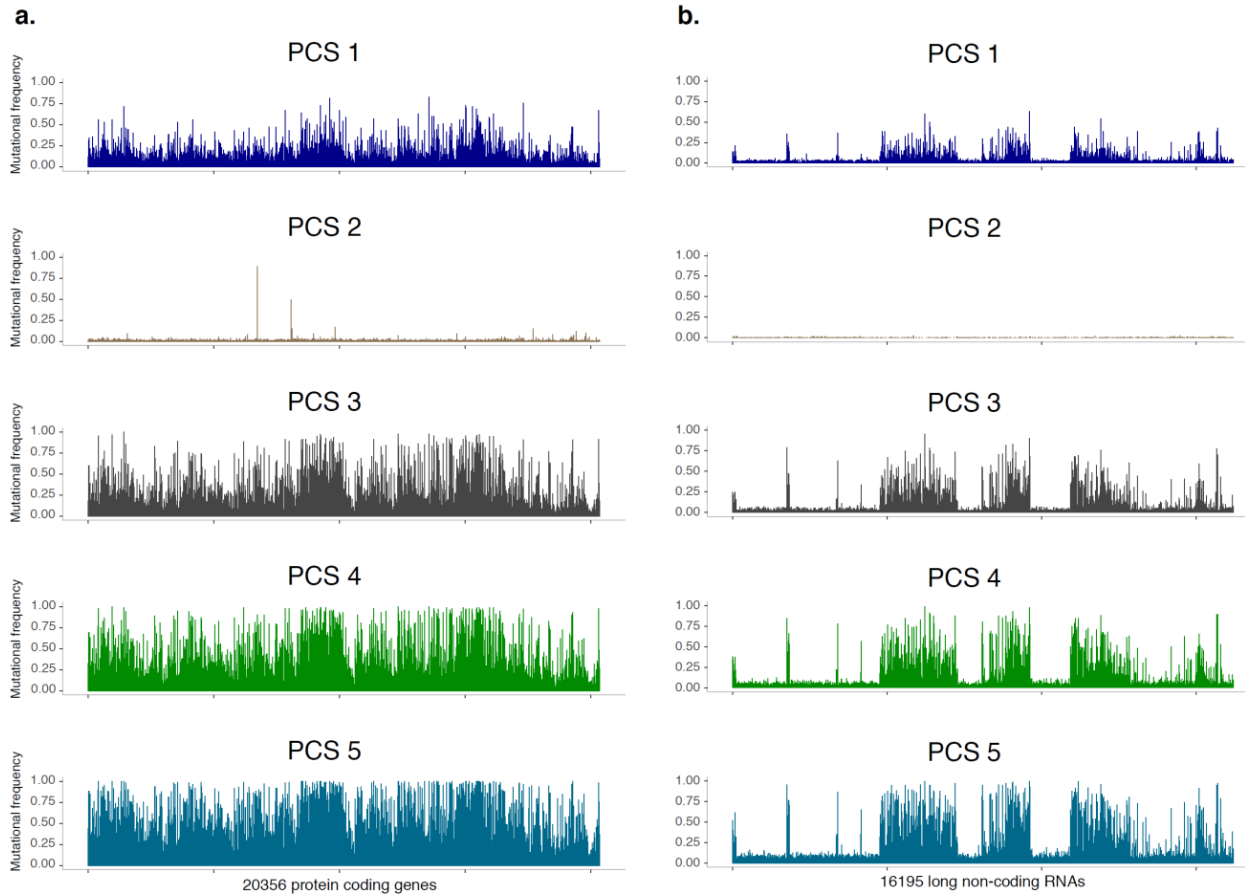

**Figure 4. Mutational load of coding and non-coding genes.** *a) mutational load of coding genes in the five subtypes. b) mutational load of non-coding genes in the five subtypes.*

### *Mutational signature*

In this section we computed mutational signature for each subtype specifically. We used CANCERSIGN [33] to identify these signatures. As it is shown in Fig. 13, in the signature T>A is significantly illustrated. In the SC2 we can see C>T is the significant signature. Further, in the SC3, T>A , T>C , C>A, C>T, has been indicated as significant mutation signatures. Moreover, in SC4, C>T and C>A are two significant mutated signatures of this subtype. In the last subtype, SC5, C>T and T>C are two significant mutation signatures which has been seen.

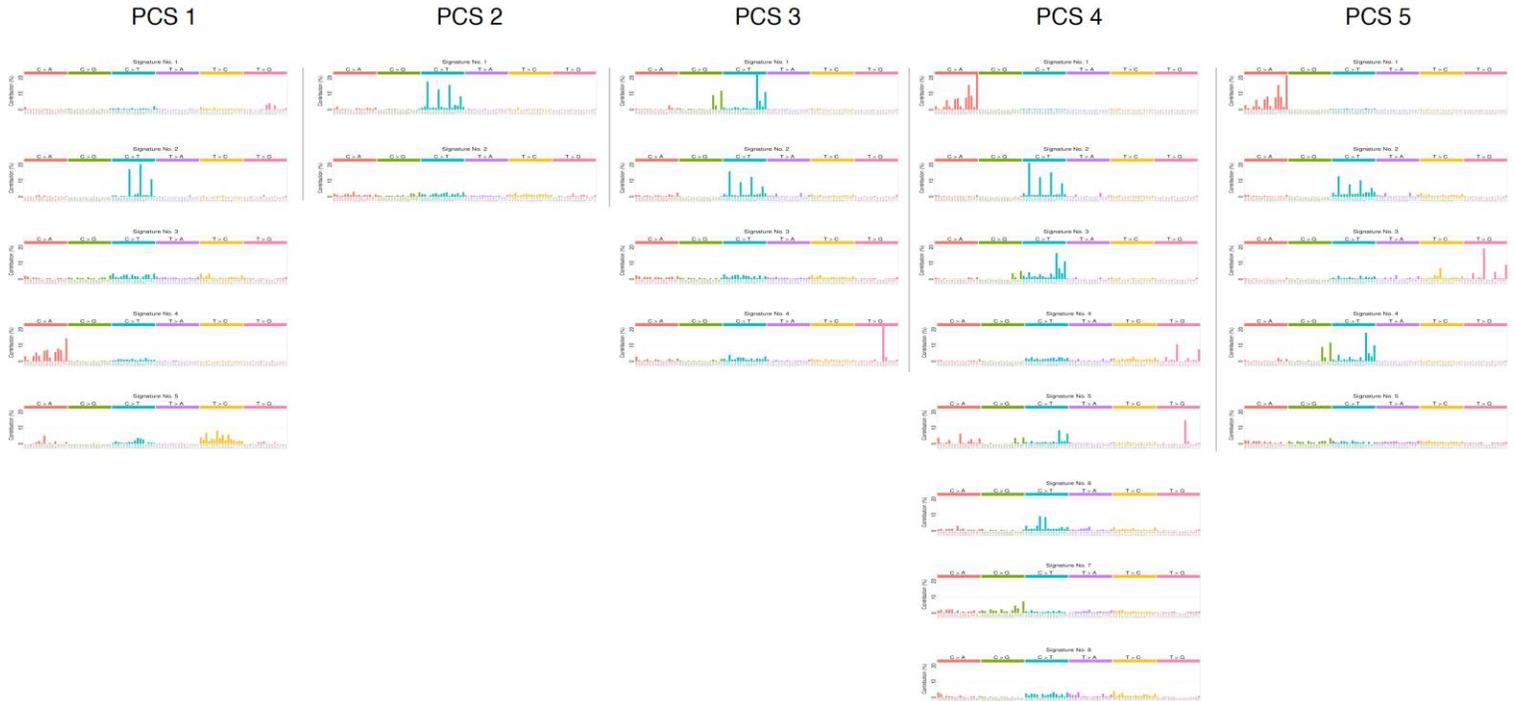

**Figure 5. Mutational analysis in the identified subtypes.** We used the CANCERSIGN tool to identify mutational signatures in each subtype. Using CANCERSIGN, we identified mutational signatures where were very similar with those identified by Alexandrov et al.

### *Motif context of mutations in each subtype*

In the motif analysis section our study was centered about just the genes which were high-mutated in each specific subtypes. We next investigated the rate of mutations in tri-nucleotides motifs of all of the genes, their proximity and differentiation with respect to the different liver cancer subtypes that are identified in this section. To do this we first computed the 3-mer motifs rate of each SSM for all of the genes (see more in the method section). As it is shown in **Figure 6**, for the SC1 subtype most of the mutational occurred in the motifs: TC-T.A, TC-A.T, TC-A.G, and TC-A.A. While the SC2 subtype comprises CT-T.C, CT-G.T, and CT-GC motifs besides those high-mutated motifs in SC1. The peak rate of mutational load in the SC3 has occurred in motifs CA-A.A, TG-G.G, TC-A.A, CT-T.T, CT-G.G, CT-C.G, CT-C.G, and CT-A.G. There are also a high-mutational load in TC-A.T, TC-A.C, and TA-CT tri-nucleotides motifs in SC4. Finally, in SC5 subtype, mutations enriched in tri-nucleotides motifs TC-A.A, TA-C.G, CT-G.G, and CT-C.C.

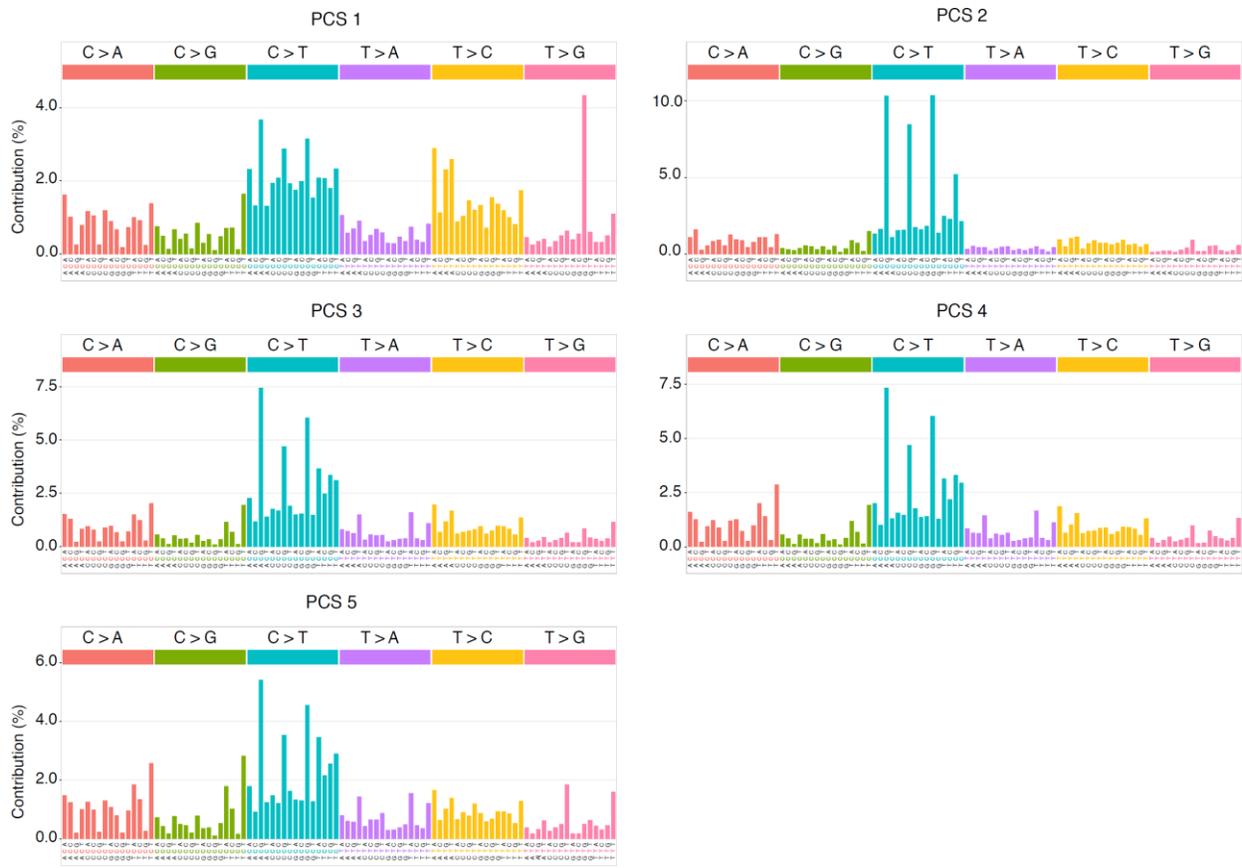

**Figure 6. Motif analysis of subtypes.** The subtypes showed different patterns for motifs indicating different mutational context.

### *Gene ontology and pathway analyses*

In this section we have obtained the results of gene ontology and pathway analyses using WEB-based Gene SeT AnaLysis Toolkit [34]. It represents gene ontology for each subtype and overall compression between all of the subtypes. For the SC1, unique gene ontologies are as the following: adherens junction organization, axonal fasciculation, cell−cell adhesion mediated by cadherin, central nervous system neuron development, forebrain generation of neurons, forebrain neuron development, forebrain neuron differentiation, heart contraction, heart process, localization within membrane, negative chemotaxis, negative regulation of cell migration, negative regulation of cell motility, negative regulation of nervous system development, negative regulation of locomotion, negative regulation of cellular component movement, neuron projection fasciculation, presynaptic membrane organization, regulation of heart rate, regulation of postsynapse organization and second−messenger−mediated signaling.

## Discussion

Despite substantial progress in liver cancer research, significant gaps remain in understanding the mutational diversity, underlying mechanisms, and clinical implications of somatic mutations in liver cancer subtypes. This study sought to address these gaps by analyzing somatic point mutations from liver cancer samples using a novel deep learning framework. By employing stacked autoencoders, we transformed mutational data from 20,356 protein-coding genes into a low-dimensional feature space while retaining the complete discriminatory information. This innovative approach allowed us to cluster samples into five distinct liver cancer subtypes (SC1 to SC5) with strong biological and statistical validation.

Our work overcomes the limitations of previous studies, which often relied on feature selection techniques that inevitably discard valuable information. By transforming all mutational data into a compressed feature set, our methodology ensures the inclusion of the entire mutational landscape, enabling a more comprehensive understanding of liver cancer heterogeneity. The identified subtypes exhibited unique mutational signatures, varying mutational loads, and distinct gene mutation profiles, underscoring the complexity of the disease and offering new opportunities for precision medicine.

The five subtypes identified in this study revealed important insights into the molecular heterogeneity of liver cancer. SC1 and SC2 displayed significantly higher mutational loads compared to SC3, SC4, and SC5, with SC3 having the lowest mutational burden. Interestingly, SC4 and SC5 demonstrated highly distinctive mutation patterns, suggesting unique molecular mechanisms driving these subtypes. This diversity highlights the necessity of analyzing liver cancer as a collection of molecularly distinct diseases rather than a single entity. For example, while SC1 and SC2 shared 73 significantly mutated genes, each also harbored 28 unique genes, emphasizing the distinct mutational landscape of these subtypes. In contrast, SC3, SC4, and SC5 exhibited even greater uniqueness, with 98 subtype-specific genes identified for each. These findings provide a rich source of potential biomarkers and therapeutic targets that could guide the development of more personalized treatment strategies.

Our mutational signature analysis offered further insights into the molecular processes driving these subtypes. The predominant C>T mutations in SC2 and the combination of T>A, T>C, C>A, and C>T mutations in SC3 point to distinct mutational processes active in each subtype. The identification of both well-known COSMIC signatures and novel subtype-specific signatures provides critical information about the etiology of liver cancer subtypes. For instance, SC4 and SC5 exhibited signatures similar to SBS10a and SBS31, which are associated with hypermutator phenotypes and chemotherapy-induced mutations, respectively. Such insights into mutational signatures could inform the development of targeted therapies and improve our understanding of drug resistance mechanisms. Additionally, the unique mutational patterns observed across subtypes suggest the presence of subtype-specific molecular pathways that could be exploited for therapeutic intervention.

Our motif context analysis further refined our understanding of mutational differences between subtypes. By examining trinucleotide mutation motifs, we identified subtype-specific patterns that provide additional clues to the molecular mechanisms driving liver cancer. For example, SC1 mutations were enriched in motifs like TC-T.A and TC-A.T, while SC4 exhibited a distinct enrichment in TC-A.C and TA-CT motifs. These findings underscore the importance of considering not only the frequency of mutations but also their context, as these patterns may offer new avenues for biomarker discovery and drug development.

Gene ontology and pathway analyses provided functional validation of our findings. SC1 was enriched for pathways related to central nervous system neuron development, regulation of heart rate, and axonal fasciculation, while SC3 showed unique enrichment in pathways associated with negative regulation of cell migration and motility. These distinct functional profiles align with the mutational differences observed across subtypes and suggest that each subtype is associated with specific biological processes. Such insights could guide the development of subtype-specific therapeutic strategies, further advancing precision medicine in liver cancer.

Our findings also have important clinical implications. By integrating somatic mutation data with advanced clustering techniques, we have provided a robust framework for identifying molecularly distinct liver cancer subtypes. This approach not only improves our understanding of the disease's complexity but also offers a foundation for developing personalized treatment plans. For example, the identification of subtype-specific gene-motif profiles could help clinicians design targeted therapies that account for the unique mutational context of each subtype. Moreover, the association of mutational signatures with known etiological factors, such as DNA repair deficiencies and chemotherapy resistance, provides critical insights into the mechanisms driving treatment response and resistance.

In addition to improving therapeutic strategies, our work highlights the potential for leveraging mutational data to develop biomarkers for early detection and prognosis. The unique gene-motif profiles and mutational signatures identified in this study could serve as the basis for diagnostic tools that enable more accurate classification of liver cancer subtypes. Furthermore, the integration of clinical information, such as survival rates and geographic or demographic factors, could refine our understanding of the interplay between genetic, environmental, and lifestyle factors in liver cancer progression.

Finally, our study demonstrates the power of deep learning in high-throughput genomics research. By transforming the high-dimensional mutational landscape into a low-dimensional feature space, we were able to overcome computational challenges and extract meaningful biological insights. This approach represents a significant step forward in the application of machine learning to cancer research and sets the stage for future studies that aim to integrate multi-omics data for a more comprehensive understanding of cancer biology.

In conclusion, our study provides a detailed characterization of liver cancer subtypes based on somatic mutations, highlighting the value of deep learning for uncovering the molecular heterogeneity of cancer. By identifying five molecularly distinct subtypes with unique mutational profiles, signatures, and functional pathways, we have laid the groundwork for advancing precision medicine in liver cancer. These findings not only enhance our understanding of the disease's complexity but also offer a roadmap for developing personalized treatment strategies, diagnostic tools, and therapeutic interventions tailored to the unique characteristics of each subtype. Future research should aim to validate these findings in larger and more diverse cohorts and explore their implications for clinical practice.

## Methods and Martials

### Dataset and Study Design

In the present work, we have used ICGC Simple Somatic Mutation (SSM) from individuals with Liver cancer. In the ICGC, there are 1,139 samples with liver cancer, including 96% whole genome sequencing and the rest of them whole exome sequencing samples. We also used Ensemble gene list to SSMs to protein-coding genes [35]. As a result, we build a benchmark consisting of 1,139 liver samples, which consist20, 159 protein-coding genes.

### Pre-processing

Our employed benchmark of 1,139 samples contain ICGC samples ID and the positions of the simple somatic mutations. We only considered simple somatic mutations and insertions and deletions are excluded from the analyses. In order to apply our method to the benchmark which contained all of the SSMs, we constructed a matrix which has one sample in each row and for each sample, we have counted the number of the SSM for each protein-coding gene. Consequently, our benchmark has a matrix with a dimension of 1,139*20,159 in which 20,159 is the total number of coding genes according to the Ensemble annotation.

### Deep learning

Recently, deep learning attracted tremendous attention in a wide range of applications in medical research including medical image and signal analyses [36-40], cancer research and bioinformatics [20, 41, 42]. In general, deep neural network is the extension of ordinary artificial neural network with more than two hidden layers which tries to model complex concepts based on learning data representations. This procedure is implemented with the help of deep graphs and multiple processing layers in which each processing layer contains linear or non-linear layers and nodes. Deep learning can be executed supervised, unsupervised, and semi-supervised learning approaches [43-45]. Each on these categories have been used in the medical biological data analysis [46-48]. Deep neural networks usually have an input layer, an output layer and multiple hidden layers between input and output layers [49]. Here, we used stack autoencoder deep neural network for feature transformation and to map our input data into a lower dimensional feature group without compromising the information embedded in our feature space.

**Stacked Autoencoder (SAE)**

An autoencoder is considered as a Deep Learning algorithm aiming at mapping the input feature space to a lower dimensional feature group [54]. An autoencoder always comprises two major parts, encoder section and decoder section. The input and output are identical in this model and the aim is to efficiently conduct this mapping in a non-linear manner. Stacked autoencoder mainly is used for feature reduction in which the input can be transformed into a new feature group in a nonlinear manner. In the implementation procedure of the autoencoder, if the number of neurons in the encoder layer (also called code layer) is less than the number of the input layer neurons, it can be treated as a compressed description of the inputs data. Further, we have normalized our dataset between 0 and 1 and also we utilized 80% of the dataset for training section (911 samples) and we used 20% for testing part (228 samples).

**Autoencode Architecture**

We have implemented our deep learning model in python (3.7) with Keras using Tensorflow as its backend. In the implementation procedure of the autoencoder for encoder part, we utilized three hidden layers besides input layer with 20,159 neurons (number of the protein-coding genes). Batch Normalization layer as the second layer in order to balance our input data, a hidden layer with 1000 neurons as the third layer and finally we have three neurons for our encoder layer. Hence, we have a compressed representation of the input data in our layer. Additionally, in the decoder part, we used three hidden layers with three, 1000 neurons, batch normalization respectively and the output layer with 20,159 neurons just as same as the encoder hidden layers.

We can formulate encoder and decoder parts as transitions of $\alpha$ and $\varphi$ such that:

$$\alpha: \chi \rightarrow \mathcal{F} \tag{1}$$

$$\varphi: \mathcal{F} \rightarrow \chi \tag{2}$$

$$\alpha, \varphi = argmin \parallel \chi - (\alpha \circ \varphi)\chi \parallel \tag{3}$$

Here we assume that we do not have any hidden layers in our autoencoder and it is in its simplest mode, inputs $x \in \mathbb{R}^d = \chi$ are mapped into:

$$z \in \mathbb{R}^p = \mathcal{F}: \tag{4}$$

$$z = \sigma(Wx + b) \tag{5}$$

Where $z$ is our code, latent representation, or the bottleneck variable, $W$ is weight matrix, $b$ is biases, and $\sigma$ is the activation. Sigmoid and rectified linear models [REF] are two most commonly used activation functions in deep learning models. Here we have also utilized sigmoid function for the seven first layers and also used rectified linear unit function for our output layer. In the decoder part of the autoencoder $z$ maps to the $\chi'$ and the shape of $\chi'$ is as same as the shape of $\chi$ :

$$\chi' = \sigma'(W'x + b') \tag{6}$$

Here for the hidden layers $\sigma$ is the same as $\sigma'$.

To reduce the reconstruction error (also called loss) we have utilized Root Mean Squared (RMS) as our loss function:

$$\mathcal{L}(\chi, \chi') = \sqrt{\frac{\sum_{n=1}^{n}(\chi_n - \chi'_n)^2}{n}} \tag{7}$$

As it was mentioned earlier, $\mathcal{F}$ has a lower dimensionality from our inputs and it is a compressed representation of our inputs ($\alpha: \chi \rightarrow \mathcal{F}$). There are usually three steps for training an autoencoder with backpropagation. First, we execute feed-forward to compute all of the values for neurons, activation functions, weights, biases and producing $\chi'$ in our output layer for each input $\chi$ in our input layer. In the second step, we compute the deviation of the $\chi'$ in our output from $\chi$ in input with measurement Root Mean Squared Error (RMSE). Finally, we backpropagate the computed error through our neural network in order to update or weights in links between neurons. In this study, we have utilized stochastic gradient decent method and specifically "Adam" optimizer for our training stage of our autoencoder model.

After training our deep learning model, we extract features (compressed representation) from our code layer. For this task we have entered all of our dataset 1139 samples into our trained model for feature extraction. Here, we have extracted 3, 5, and 100 features for each sample in which for all the cases we obtain similar RMSE. Hence, we use 3 as the number of features extracted from all our input data to obtain the lowest dimensionality. As a result, we build a feature matrix with dimension of (1139 * 3).

**Clustering**

After extracting features for each 1,139 samples of our benchmark, we clustered these samples with K-means method. To find optimum number of the clusters we used a R package "NbClust" [32]. NbClust comprises almost 27 different statistical methods (called indices in this package), which assay the clusters scores regarding each method. After computing all of the scores, NbClust indicates that 5 is the optimum number of clusters in our data using K-means method. As it is shown in **Figure 1**, in the first clustering step, we determined two clusters as an optimum number of clusters. Further, in the second step of clustering, regarding NbClust statistical scores, we determined two clusters driven from the primary first cluster and three cluster which have been drive from primary second cluster.

**Motif context**

We counted the number of mutations in 96 Tri-nucleotide motifs for each subtype and then normalized those values by dividing them to the number of samples in each cluster. introduced those with over 2% mutations occurrence in samples in each cluster. The frequency of 96tri-nucleotide motifs in each subtype is provided in **Supplementary table X**

*Motifs and mutational signature analysis*

Here we also specified tri-nucleotides motifs of each simple somatic mutation using three R packages namely, VariantAnnotation, GenomicRanges, and BSgenome.Hsapiens.UCSC.hg19 [23, 50, 51] in order to exploit them in tri-nucleotides motif analysis as well as signatures analysis sections.

*Finding most associated genes in each subtype*

Using Fisher's exact test, we have obtained the most associated genes in each subtype in terms of their mutational rate. In the present work we have utilized this test with a table for all of the 20159 protein-coding genes and calculate p-value for each one of them with respect to a given cluster. Subsequently, we sorted all of the genes with their respected p-value and selected the first 100 genes for each cluster. Hence, in this way we found the most associated genes regarding their mutational rate in each subtype. Further, we plotted these genes in the Venn diagram with R package "VennDiagram", in order to show their possible overlaps.

*Consequence type of mutations*

As consequence type of mutations were recorded in ICGC dataset, therefore, we used those information to identify consequence type of mutations in our subtypes. As there were different numbers of consequence type for each mutation, in order to prevent any biase, we have normalized total number of each consequence type by dividing them to the total amount of unique mutations.

### *Gene ontology analysis, gene pathway analyses on the significantly mutated coding genes*

In the present study we have investigated the GO for the top 100 genes, which are significantly associated to each subtype regarding their P-value. We also utilized WEB-based Gene SeT AnaLysis Toolkit [34] to investigate gene ontology and pathwayanalysis of the genesMoreover, we set  the parameters of this tool as follows: our organism of Interest was "hsapiens", method of interest was "overrepresentation enrichment analysis (ORA)", andour functional database was "geneontology and biological process". Further, as we mentioned earlier our gene ID type was "Ensemble gene ID" and reference set for enrichment analysis was "genome protein-coding". For pathway analysis parameters were set as follows: organism of Interest and method of interest was "hsapiens", our functional database was "pathway and KEGG"; also our gene ID type was "Ensemble", and reference set for enrichment analysis was "genome protein-coding". In the both analyses, we filtered out those associations with P- > 0.05.

## Conflict of interest

The authors declare no competing financial and non-financial interests.